\documentclass{article}

\usepackage{arxiv}

\usepackage[utf8]{inputenc} 
\usepackage[T1]{fontenc}    
\usepackage{hyperref}       
\usepackage{url}            
\usepackage{booktabs}       
\usepackage{amsfonts}       
\usepackage{nicefrac}       
\usepackage{microtype}      
\usepackage{lipsum}
\usepackage{graphicx}
\graphicspath{ {./images/} }
\usepackage{lmodern} 
\usepackage{microtype}
\usepackage[numbers,square,sort&compress]{natbib}
\usepackage[toc,page]{appendix} 
\usepackage{lipsum, amsmath}

\title{A Win-Expectancy Framework for Contextualizing Runs Batted In: Introducing ARBI and CRBI}

\author{
 Wuhuan Deng \\
  Department of Applied Mathematics\\
  University of Washington\\
  Seattle, WA 98105 \\
  \texttt{wudeng@uw.edu} \\
}

\begin{document}
\maketitle	

\begin{abstract}
    Runs Batted In (RBI) records the number of runs a hitter directly drives in during their plate appearances and reflects a batter’s ability to convert opportunities into scoring. Because producing runs determines game outcomes, RBI has long served as a central statistic in evaluating offensive performance. However, traditional RBI treats all batted-in runs equally and ignores the game context in which they occur, such as leverage, score state, and the actual impact of a run on a team’s chance of winning. In this paper, we introduce two new context-aware metrics—Adjusted RBI (ARBI) and Contextual RBI (CRBI)—that address the fundamental limitations of RBI by incorporating Win Expectancy (WE). ARBI rescales each RBI according to the change in WE before and after the scoring event, assigning more value to runs that meaningfully shift the likelihood of winning and less to runs scored in low-leverage situations. We then extend this framework to CRBI, which further differentiates RBIs with the same WE change by accounting for the terminal WE at the end of the event. This refinement captures the idea that an RBI increasing WE from, for example, 0.45 to 0.65 has a larger competitive impact than one increasing WE from 0.05 to 0.25, even though both represent a 20\% increase. Together, ARBI and CRBI provide calibrated, context-sensitive measures of offensive contribution that more accurately reflect the true value of run production. These metrics modernize the interpretation of RBI and have broad applications in player evaluation, forecasting, contract valuation, and decision-making in baseball analytics.
\end{abstract}

	
\section{Introduction} 
\subsection{Motivation}
\begin{figure}
  \centering
  \includegraphics[width=0.48\textwidth]{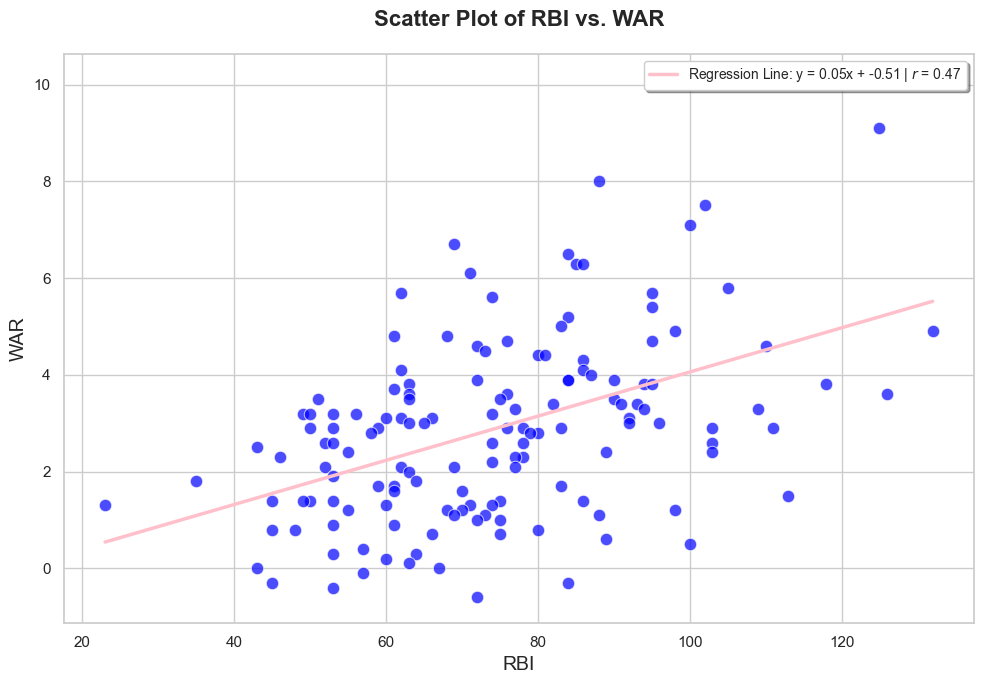}
  \includegraphics[width=0.48\textwidth]{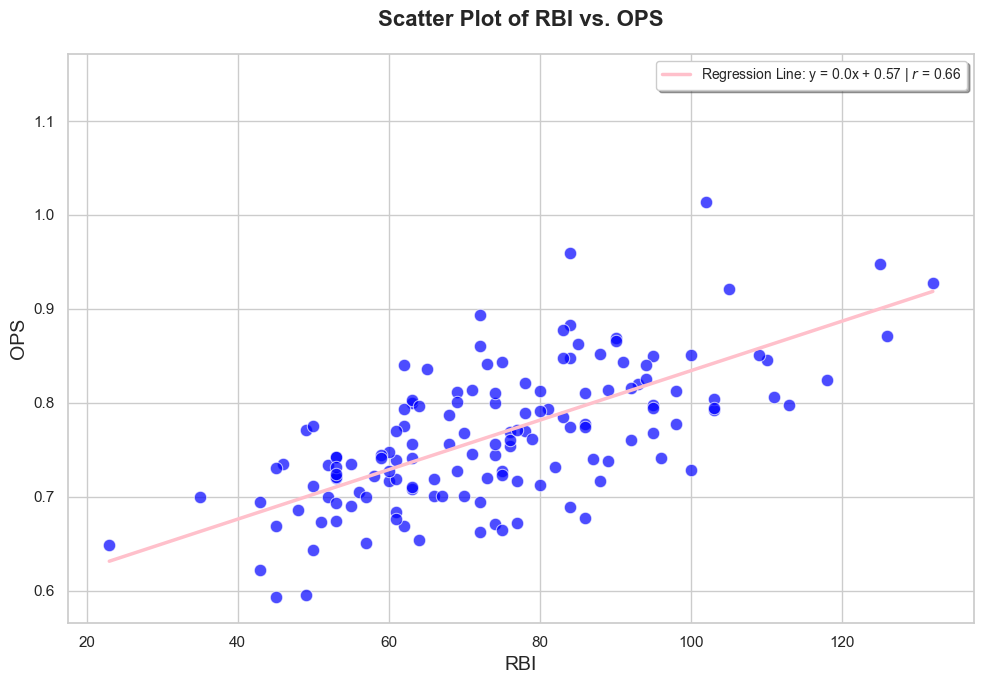}
  \caption{Relation between RBI and WAR/OPS in 2025 MLB regular season.}
  \label{RBI-WAR}
\end{figure}

RBI has long been one of the most recognized statistics in baseball. It directly captures a batter's ability to drive in runs, which lead to winning game. And it has played a central role in player evaluation, contract negotiations, award voting, and Hall of Fame debates. An example is the batting Triple Crown, one of most prestigious achievements, which is defined jointly with batting average, home runs, and RBI. This suggests that how deeply RBI is embedded in both the culture and recognition in baseball. From Figure~\ref{RBI-WAR}, we filter players with at least 50 games played in MLb 2025 season and we can see that RBI has positive relation with both OPS and WAR, which are also two important metrics in evaluating batters.

However, despite the importance of RBI, it has several limitations. Cyril Morong has used regression method to show that RBI has heavily depends on opportunities \cite{chance} and Neil Weinberg has emphasized that RBI is not just an individual skill measure \cite{team:rbi}. In addition, RBI treats all runs equally regardless of game situation. These suggest that while RBI captures a meaningful aspect of offensive performance, it is still not direct enough to evaluate hitters batted in runs contribution to win the game. For example, a one run homer when the score is 0-0 at bottom of 8th inning is much more valuable than a one run homer when the home team is leading 5-0 at bottom of 8th inning, but current RBI just record this as 1 RBI and do not reflect the true contribution to the win.

Therefore, we introduce ARBI and CRBI by rescaling each RBI to reflect how much this RBI contributes in winning. We define ARBI based on the change of WE, where the increase of WE will lead to a larger RBI. However, even with same change of WE, each RBI is also different under different context. For example, there are two RBIs and they also increase the team WE by 20\%. In contrast, one RBI increases the WE from 0.46 to 0.65, but another one increases the WE from 0.05 to 0.25. The first RBI should deserve more value than the later one, because that RBI leads the team from underdog to favorite, but the second RBI although increasing 20\% WE, the team is still very likely to lose the game. Therefore, we further rescale ARBI under each same change of WE by considering the terminal WE after the event.

\subsection{Related Works} 
Although there are already existing works related to RBI, those works have largely focused on correcting the metric;s dependence on opportunity and team context. Prior analysis show that RBI totals are heavily by baserunners, lineup position, and overall run environment, leading to some adjusted metrics. Neil Weinberg introduced RE24 in 2014, which measure how many runs a team is expected to score from each of the 24 possible base-out states. By comparing the expected runs before and after a plate appearance, RE24 captures the true run value of an event in its specific game context \cite{RE24}. FanGraphs has also introduced Context-Neutral RBI (cnRBI), a metric that estimates how many RBI a hitter should produce under league-average opportunities and base-out states, effectively removing the influence of lineup position and teammate quality \cite{cnRBI}. However, these approaches primarily address opportunity bias and run value, instead of the actual impact of a scoring event on game outcomes. In contrast, our proposed ARBI and CRBI metrics incorporate WE change and base state, which provide a fully score context-sensitive evaluation of run production that directly ties each RBI to its effect on winning.

\section{Methods}
\subsection{Win Expectancy (WE)}
\label{sec:WE}
Win Expectancy (WE) is the percent chance a particular team will win based on the score, inning, outs, runners on base, and the run environment \cite{WE}. The values are computed based on historical data. A famous WE table is created by Tom Tango's, which considers inning, outs, base runners, and score difference and is based on 2010 - 2015 MLB game data \cite{WE Table}. A simple WE table example is shown in Table~\ref{WE-Tab-ex}. All the values in from Tom Tango's table are respect to home team. For example, when the home team is trailing by 5 points at bottom of 1st inning with 0 outs and empty bases, the WE of the home team is 0.128. 
\begin{table}
  \centering
  \begin{tabular}{c c c c c c c c c c c c c c c}
    \hline
    Inn & Top{/}Bot & Outs & Runners & -5 & -4 & -3 & -2 & -1 & Tie & +1 & +2 & +3 & +4 & +5 \\
    \hline
    1 & Bottom & 0 & Empty & 0.128 & 0.184 & 0.255 & 0.342 & 0.442 & 0.547 & 0.649 & 0.739 & 0.814 & 0.871 & 0.914 \\
    1 & Bottom & 0 & 1B Only & 0.153 & 0.214 & 0.291 & 0.381 & 0.480 & 0.583 & 0.679 & 0.764 & 0.832 & 0.885 & 0.923 \\
    \hline
  \end{tabular}
  \caption{Example of Tom Tango's WE table of bottom of 1st inning with 0 outs, when bases are empty or runner on 1st base.}
  \label{WE-Tab-ex}
\end{table}
In this research, all WE will be computed only based inning, top or bottom, outs, base runners, and score difference and we only consider regular 9 innings. When score difference is between -5 and +5, we will directly use the value from Tom Tango's Table, and for values outside of the empirical range (theoretically, in a baseball game, score difference $ \in (-\infty, +\infty) $), we are going to use spline method to evaluate WE on its entire domain. For more details of this computation, see Appendix~\ref{extend WE}. Figure~\ref{fig:WE extend} shows the values of WE extended to score difference from -10 to 10 at the bottom of 6th inning with 0 outs.
\begin{figure}
  \centering
  \includegraphics[width=\textwidth]{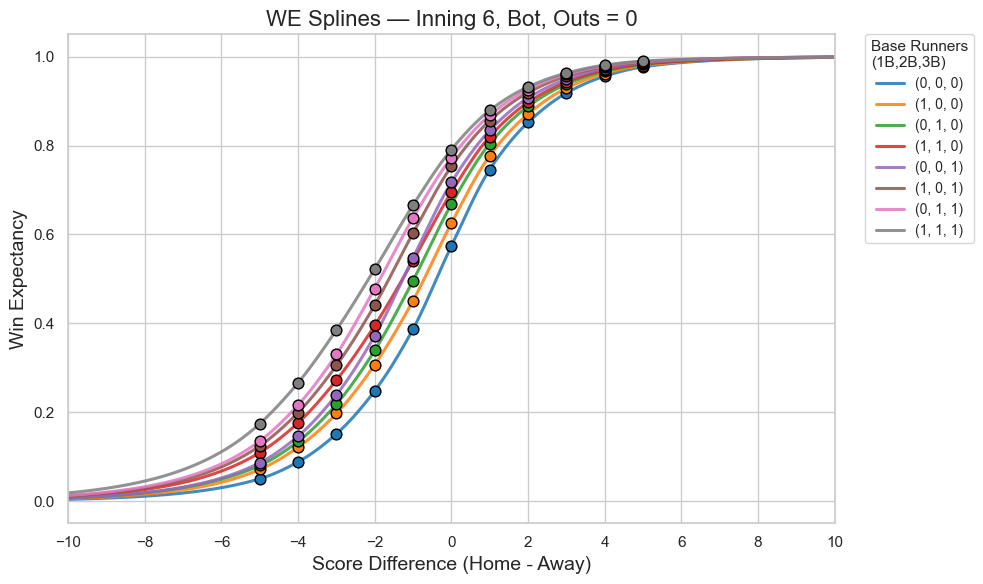}
  \caption{Example of WE extended to the entire domain under certain game state. The dots are the empirical value given by Tom Tango's table.}
  \label{fig:WE extend}
\end{figure}

\subsection{Runs Batted In (RBI)}
A batter is credited with an RBI in most cases where the result of his plate appearance is a run being scored. There are a few exceptions, however. A player does not receive an RBI when the run scores as a result of an error or ground into double play. The most common examples of RBIs are run-scoring hits. However, players also receive an RBI for a bases-loaded walk or hit by pitch. Players can earn RBIs when they make outs, as well, provided the out results in a run or runs (except, as noted above, in the case of double plays)~\cite{MLB RBI}. However, in this research, we have a slightly different way to define RBI. Since we believe that any run scoring is related to the batter during a plate appearance no matter it happens because of a wild pitch or an error. Therefore, in this research, the hitter is awarded RBIs whenever the run scores during or after his plate appearance. Since these extra situations are rare, so we will only have a subtle difference between our RBI and the actual RBI number. In 2025 MLB regular season, based on FanGraphs, with at least 30 games played, the mean value of accumulated RBI is 74.37 and the mean value of our accumulated RBI is 76.187.

\subsubsection{Define Adjused RBI (ARBI)}
Adjusted RBI is rescaled by the change of WE when run scores. It captures the true contribution of each RBI in changing the win expectancy. It is defined as:
\begin{equation*}
  \text{ARBI} = \alpha \cdot \text{RBI},
\end{equation*}
where adjustment factor $ \alpha $ is determined by the change in win expectancy associated with scoring event. A single event may produce multiple RBIs, so we assign the same $ \alpha $ value to each run driven in during that event. Let $ \Delta \text{WE} = \text{WE}_{start} - \text{WE}_{end} $, where $ \text{WE}_{start} \text{ and } \text{WE}_{end} $ represent the WE before and after the scoring event. And the way of evaluating WE in this research has been mentioned in Section~\ref{sec:WE}. Mathematically, $ \alpha $ is defined as:
\begin{equation*}
  \alpha = f(\Delta \text{WE}), \Delta \text{WE} \in (-1, 1),
\end{equation*}
where $ f $ is a monotonically increasing function. Since the WE is not just defined on score difference but also game state, so even with a run score, the WE could also decrease. This is why $ \Delta $WE could be less than 0. For example, when the home team is trailing by 3 scores at bottom of 9th inning with 1 out and runner on 3rd base. Then the hitter hits a sacrifice play where the hitter is out, but the runner on third base scores. Hence, the game state becomes the home team is trailing by 2 scores at bottom of 9th inning with 2 outs and empty bases. In this example, the $ \text{WE}_{start} = 0.051 $ and $ \text{WE}_{end} = 0.014 \Rightarrow \Delta \text{WE} = -0.037 $. The domain of ARBI is defined on (0, 2) so that after rescaling, the original one RBI will not disappear even it does not influence the game at all and will also not surpass two RBIs no matter how important it is. Therefore, the domain of $ \alpha $ is also (0, 2). Hence, based on the domain of $ \Delta \text{WE} $ and $ \alpha $, when $ \Delta \text{WE} $ is approaching to -1 or 1, the $ \alpha $ will approach to 0 or 2, but never reach it. By this method, when the RBI increases WE more, it will be awarded more compared with another RBI which increases WE less.

\subsubsection{Define Contextual RBI (CRBI)}
The limitation of ARBI is obvious. ARBI treat each RBI with same $ \Delta \text{WE} $ equally. An example has been mentioned above. For two RBIs with $ \Delta \text{WE} = 0.2 $, one RBI increases the WE from 0.46 to 0.65, but another one increases the WE from 0.05 to 0.25. The first RBI should deserve more value than the later one, because that RBI leads the team from underdog to favorite, but the second RBI although increasing 20\% WE, the team is still very likely to lose the game. Therefore, to ensure each RBI could be reflected to its true value and contribution to the win, we introduce CRBI which further rescales ARBI based on $ \text{WE}_{end} $, and it is defined as:
\begin{equation*}
  \text{CRBI} = \beta \cdot \text{ARBI} = \beta \cdot \alpha \cdot \text{RBI},
\end{equation*}
where $ \beta = g(\Delta \text{WE}, \text{WE}_{end}) $. The domain of CRBI is also (0, 2) which is same with ARBI so that we ensure each RBI will not disappear or double. Since CRBI is based on ARBI, so $ \beta \in (0, \frac{2}{\alpha}) $.

\subsection{Estimate $ f $}
\begin{figure}
  \centering
  \includegraphics[width=\textwidth]{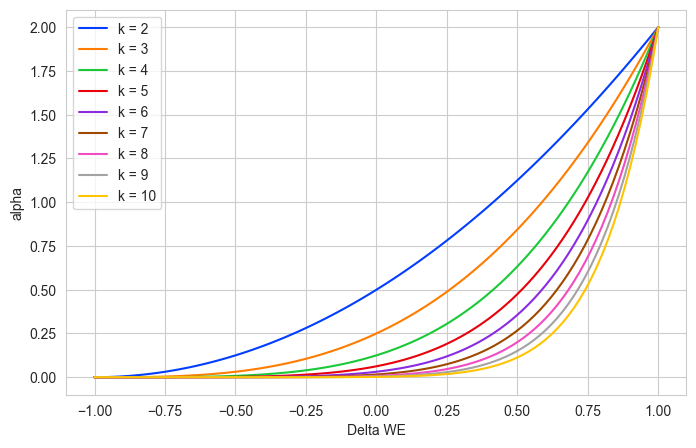}
  \caption{Example of $ \alpha $ of different choices of $ k $ based on power function.}
  \label{fig:power f}
\end{figure}

\begin{figure}
  \centering
  \includegraphics[width=\textwidth]{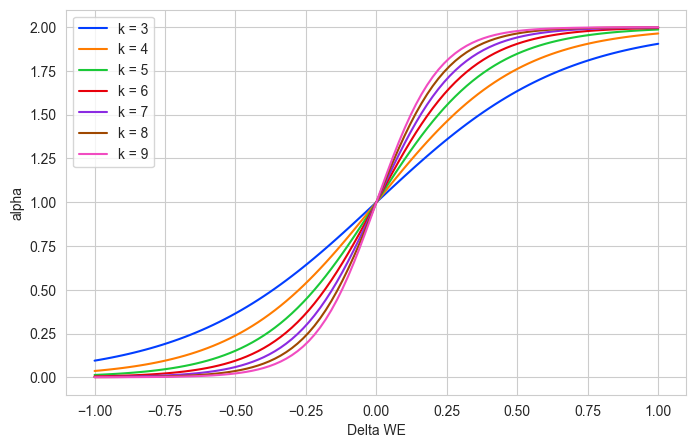}
  \caption{Example of $ \alpha $ of different choices of $ k $ based on sigmoid function.}
  \label{fig:sigmoid f}
\end{figure}

It is essential to emphasize that there should not be a best or correct way to determine $ f $. Since the goal of ARBI is to rescale each RBI based on the change of WE, $ f $ could be determined in different ways based on the actual use scenarios as long as: 1). $ f $ is monotone increasing, and 2). $ \alpha = f(\Delta \text{WE}) \in (0, 2) $. For example, $ f $ could be a linear function so that each change of $ \Delta \text{WE} $ will be treated same, or $ f $ could be a convex function, which means that when $ \Delta \text{WE} $ is getting larger, it will be awarded more. Previously, we have claim that when $ \Delta \text{WE} $ is approaching to -1 or 1, the $ \alpha $ will approach to 0 or 2, but never reach it, but when estimating $ f $, we can set $ f(-1) = 0 $ and $ f(1) = 2 $ for convenient computation. In Figure~\ref{fig:power f}, we let $ \alpha = f(\Delta \text{WE}) = 2 (\frac{\Delta \text{WE} + 1}{2})^k $, where $ k $ controls the shape of the curve to determine how much change in WE will result how much change in $ \alpha $. In Figure~\ref{fig:sigmoid f}, we let $ \alpha = f(\Delta \text{WE}) = \frac{2}{1 + e^{-k \Delta \text{WE}}} $, which is based on sigmoid function. This allows $ \alpha $ to increase slow when $ \Delta \text{WE} $ is less than 0 and increase fast when $ \Delta \text{WE} $ is larger than 0 and slow again when $ \Delta \text{WE} $ is approaching to 2.

\subsection{Estimate $ g $}
\begin{figure}
  \centering
  \includegraphics[width=\textwidth]{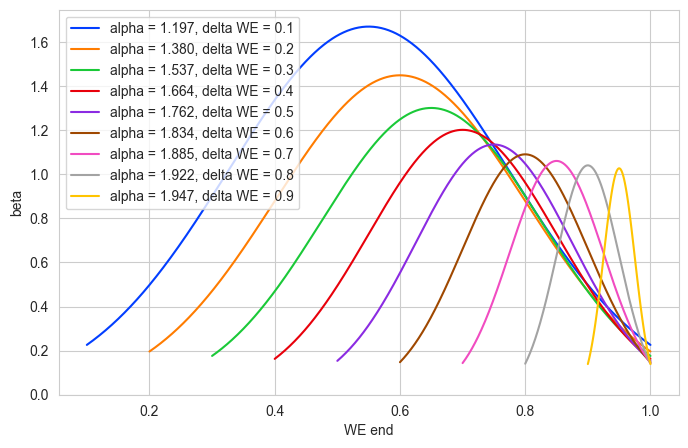}
  \caption{Example of $ \beta $ under different values of $ \alpha $, with selected function $ f $.}
  \label{fig:beta}
\end{figure}

$ \beta $ rescales ARBI under same $ \alpha $ or $ \Delta \text{WE} $ based on the actual win expectancy after score runs. Therefore, we need to define at which point, the RBI should be credited most and least. In general, for a same value of $ \Delta \text{WE} $, if the $ \Delta \text{WE} $ does not really influence the final result, it should deserve less value. Hence, for a specific value of $ \Delta \text{WE} $, we obtain $ \max(\beta) $ at $ \text{WE}_{end} = \frac{1 + \Delta \text{WE}}{2} $, which is the middle value of $ \text{WE}_{end} $ domain for given $ \Delta \text{WE} $, and $ \beta $ should follow a bell shaped curve for each given $ \Delta \text{WE} $. And when $ \Delta \text{WE} <= 0 $, ARBI will already be very small or even close to 0, so we will let $ \beta $ automatically be 1. Hence, we only need to estimate the $ g $ when $ \Delta \text{WE} > 0 $. We define:
\begin{equation*}
  \beta = g(\Delta \text{WE}, \text{WE}_{end}) = \frac{2}{\alpha} \cdot e^{-\frac{(\text{WE}_{end} - \mu)^2}{2 \sigma^2}},
\end{equation*} where $ \alpha = f(\Delta \text{WE}) $, $ \mu = \frac{1 + \Delta \text{WE}}{2} $, and $ \sigma = \min(\frac{\mu - \Delta \text{WE}}{2}, \frac{1 - \mu}{2}) $. By this method, $ \beta $ will follow a normal distribution shape for each fixed $ \alpha $. In Figure~\ref{fig:beta}, we show a simple example of estimating $ \beta $, where we choose $ f(\Delta \text{WE}) = \frac{2}{1 + e^{-4 \Delta \text{WE}}} $. Then based on this chosen $ f $, we choose 9 different values of $ \Delta \text{WE} $, and plot out the curve of $ \beta $. We can see that when $ \Delta \text{WE} $ is smaller, $ \beta $ rescales more. This is reasonable, because when $ \Delta \text{WE} $ is very large, $ \text{WE}_{end} $ only belongs to a very small of domain, and when $ \Delta \text{WE} $ is large which means that the RBI should be credited a lot.

\section{Results}
\begin{figure}
  \centering
  \includegraphics[width=\textwidth]{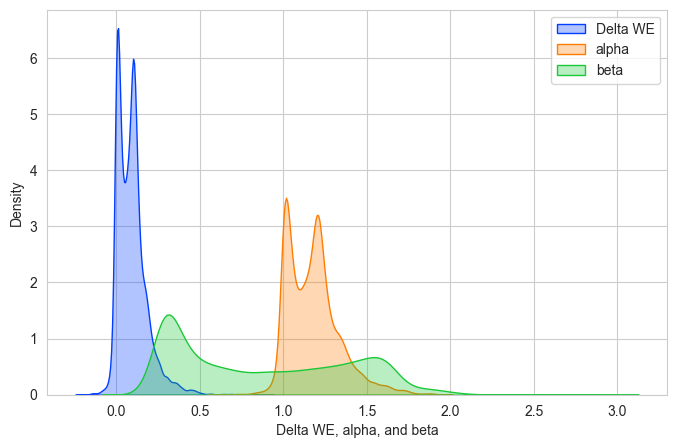}
  \caption{Distribution of $ \Delta \text{WE}, \alpha, \text{ and } \beta $ of 2025 selected score events in regular 9 innings.}
  \label{fig:metrics dist}
\end{figure}

We choose any score event from MLB 2025 regular season in regular 9 innings and compute their ARBI and CRBI with chosen function $ f(\Delta \text{WE}) = \frac{2}{1 + e^{-4 \Delta \text{WE}}} $, and the corresponding function $ g $. There are 16182 score events in total. The mean of $ \Delta \text{WE} $ is 0.093 and standard deviation is 0.091. The corresponding mean value of $ \alpha $ and $ \beta $ is 1.177 and 0.878. In Figure~\ref{fig:metrics dist}, we plot out the distribution of $ \Delta \text{WE}, \alpha, \text{ and } \beta $.

\begin{figure}
  \centering
  \includegraphics[width=\textwidth]{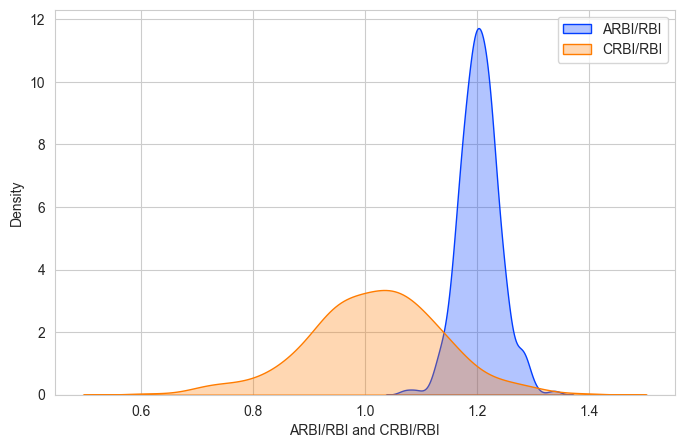}
  \caption{Distribution of ARBI/RBI and CRBI/RBI}
  \label{fig:dist /}
\end{figure}

In Table~\ref{tab:top10RBI}, we show 10 hitters with most RBI through out the entire 2025 regular season based on our computation rule. And we show these hitters' ARBI, ARBI/RBI, CRBI, and CRBI/RBI. If ARBI/RBI and CRBI/RBI are larger, it means hitters generate RBI more under critical situation and contribute more in actual wins compared with same amount of RBI. In the following analysis, we only select hitters with at least 30 RBI in the 2025 regular season. In Figure~\ref{fig:dist /}, we plot out the distribution of ARBI/RBI and CRBI/RBI and in Table~\ref{tab:change percentile}, we show the 25\%, 50\%, and 75\% percentile of ARBI/RBI and CRBI/RBI. Generally, if a hitter can obtain ARBI/RBI larger than 1.2 and CRBI/RBI larger than 1.02, the hitter could be recognized as a major contributor in actually winning the game. Table~\ref{tab:5 change} shows 5 batters with largest value of ARBI/RBI and CRBI/RBI. These players should be regarded as better players under critical conditions and deserve larger value of contract.

\begin{table}
  \centering
  \begin{tabular}{c c c c c c}
    \hline
    Batter & RBI & Adjusted RBI & ARBI/RBI & Contextual RBI & CRBI/RBI \\
    \hline
    Kyle Schwarber & 131 & 159.63 & 1.22 & 127.46 & 0.97 \\
    Cal Raleigh & 129 & 160.67 & 1.25 & 123.84 & 0.96 \\
    Pete Alonso & 124 & 153.95 & 1.24 & 138.31 & 1.11 \\
    Eugenio Suarez & 119 & 152.62 & 1.28 & 133.31 & 1.12 \\
    Vinnie Pasquantino & 116 & 141.75 & 1.22 & 127.42 & 1.10 \\
    Aaron Judge & 114 & 134.16 & 1.18 & 113.96 & 1.00 \\
    Riley Greene & 113 & 136.38 & 1.21 & 99.83 & 0.88 \\
    Junior Caminero & 109 & 129.49 & 1.19 & 96.14 & 0.88 \\
    Rafael Devers & 108 & 129.37 & 1.20 & 106.26 & 0.98 \\
    Juan Soto & 108 & 129.61 & 1.20 & 116.05 & 1.07 \\
    \hline
  \end{tabular}
  \caption{Hitters with 10 RBI and their ARBI, ARBI/RBI, CRBI, and CRBI/RBI values through out the entire season.}
  \label{tab:top10RBI}
\end{table}

\begin{table}
  \centering
  \begin{tabular}{c c c}
    \hline
    Percentile & ARBI/RBI & CRBI/RBI \\
    \hline
    25\% & 1.18 & 0.94 \\
    50\% & 1.20 & 1.02 \\
    75\% & 1.22 & 1.09 \\
    \hline
  \end{tabular}
  \caption{Percentile of ARBI/RBI and CRBI/RBI of players with at least 30 RBI.}
  \label{tab:change percentile}
\end{table}

\begin{table}
  \centering
  \begin{tabular}{c c c c}
    \hline
    Batter & ARBI/RBI & Batter & CRBI/RBI \\
    \hline
    Sean Murphy & 1.33 & Josh lowe & 1.39 \\
    Colton Cowser & 1.30 & Bo Naylor & 1.31 \\
    Colson Montgomery & 1.29 & Jose Ramirez & 1.31 \\
    Daniel Schneemann & 1.29 & Zach Neto & 1.29 \\
    Trent Grisham & 1.29 & Willy Adames & 1.27 \\
    \hline
  \end{tabular}
  \caption{5 batters with largest ARBI/RBI and CRBI/RBI.}
  \label{tab:5 change}
\end{table}

\section{Discussion}
The key ideas of ARBI and CRBI is to rescale each RBI based on the actual game situation and these new metrics should reflect more actual contribution of winning compared with RBI. In this research result section, we only use simplest method, sigmoid function to estimate $ f $. However, in the real application, based on the interval of ARBI needed (in special cases, teams or analyst can extend the domain of ARBI to make analysis more aggressive), the smoothness of $ f $, and other considerations, $ f $ could be estimated in multiple ways including regression methods, combining league average and variance, clustering, or even simple linear function. Also, the idea of ARBI and CRBI could extend to pitching. For example, reliever A has 3.50 ERA, and reliever B has 2.5 ERA. Generally, we agree that reliever B is a better choice. However, if we focus on each run they allowed, we may found out that reliever A allow runs frequently when team has 3 or more runs lead, but reliever B always allow runs when team has 2 or less runs lead and never allow runs when more than r runs lead. It is very feasible to transfer ARBI and CRBI idea to ERA and identify pitcher contribution in another insight.

\section{Conclusion}
In this paper, we proposed two new run-production metrics-Adjusted RBI and Contextual RBI-to overcome yje limitations of traditional RBI by incorporating the win expectancy of each game state and score difference. While RBI has historical value, it fails to consider game state, leverage, or the actual impact of a score on the likelihood of winning. ARBI addresses this limitation by rescaling each RBI based on the change in win expectancy based on game state and score difference, which guarantees that runs with larger change in win expectancy receives more weight. CRBI further refines the evaluation by incorporating the terminal win expectancy, allowing two RBIs with equal win expectancy change to be distinguished based on the team's eventual win expectancy. ARBI and CRBI provide a richer and more context-sensitive picture of offensive contribution. Hitters with same number of RBI show differences in ARBI/RBI and CRBI/RBI, reflecting their different performance under leverage and their true influence on game outcomes. Beyond hitting, the conceptual framework introduced in this research is flexible and can be extended to pitching metrics, defensive contributions, or other baseball events. Future work may explore integrating machine learning models to estimate functions $ f $ and $ g $ to incorporate uncertainty quantification, or apply the method to decision making. Overall, ARBI and CRBI represent a modernized, outcomes0aligned approach to evaluate each run score.

\begin{appendices}
  \section{Spline-Based Extension of the Win Expectancy Function}
  \label{extend WE}
  To construct a smooth, monotone, and asymptotically correct win expectancy (WE) function for all score differential values, including those outside the empirical range, we develop a continuous extension based on a shape-preserving Piecewise Cubic Hermite Interpolating Polynomial (PCHIP) and exponential tail functions. This method preserves the exact empirical probabilities on the observed grid while ensuring realistic limiting behavior as the score difference becomes very large. The construction follows the classical monotone cubic interpolation framework of Fritsch and Carlson~\cite{fritsch1980monotone}.

  \subsection{Empirical WE Data}
  For each game state, defined by inning $i \in \{1,\dots,9\}$, half-inning indicator 
  $h \in \{0,0.5\}$ (top or bottom), outs $o \in \{0,1,2\}$, and base-runner configuration 
  $b=(b_1,b_2,b_3)$ with $b_j \in \{0,1\}$, the empirical WE table provides valid data pairs
  \[
  (x_k, y_k), \qquad k=1,\dots,n,
  \]
  where $x_k$ is the observed score difference (home minus away) and 
  $y_k \in [0,1]$ is the corresponding win expectancy. The empirical domain typically satisfies
  \[
  x_1 < x_2 < \dots < x_n, \qquad x_k \in [-5,5].
  \]

  \subsection{Piecewise Cubic Hermite Interpolation on the Interior Domain}
  On the interval $[x_1,x_n]$, we define the interior WE function $f_{\mathrm{raw}}(x)$ as the monotone 
  Piecewise Cubic Hermite Interpolating Polynomial (PCHIP).  
  Let
  \[
  h_k = x_{k+1} - x_k, 
  \qquad 
  \delta_k = \frac{y_{k+1} - y_k}{h_k},
  \qquad k = 1,\dots,n-1.
  \]
  The interpolant on each subinterval $[x_k, x_{k+1}]$ is the cubic Hermite polynomial
  \[
  f_{\mathrm{raw}}(x)
  = 
  a_k (x - x_k)^3
  + b_k (x - x_k)^2
  + m_k (x - x_k)
  + y_k,
  \qquad x_k \le x \le x_{k+1},
  \]
  where $m_k = f_{\mathrm{raw}}'(x_k)$ are slopes chosen to preserve monotonicity.

  \subsubsection{Monotonicity-Preserving Slope Selection}
  Following Fritsch and Carlson~\cite{fritsch1980monotone}, slopes are defined as:
  \[
  m_k =
  \begin{cases}
  0, & \text{if } \delta_{k-1}\delta_k \le 0, \\[8pt]
  \dfrac{w_1 + w_2}{\dfrac{w_1}{\delta_{k-1}} + \dfrac{w_2}{\delta_k}},
  & \text{if } \delta_{k-1}\delta_k > 0,
  \end{cases}
  \]
  where
  \[
  w_1 = 2h_k + h_{k-1},
  \qquad
  w_2 = h_k + 2h_{k-1}.
  \]
  This weighted harmonic mean guarantees that $m_k$ lies between 
  $\delta_{k-1}$ and $\delta_k$, preventing overshoot. Endpoint slopes $m_1$ and $m_n$ follow one-sided variants of the same rule.

  Once the slopes $m_k$ and $m_{k+1}$ are defined, the remaining cubic coefficients are uniquely determined:
  \[
  a_k = \frac{m_k + m_{k+1} - 2\delta_k}{h_k^2},
  \qquad
  b_k = \frac{3\delta_k - 2m_k - m_{k+1}}{h_k}.
  \]
  Thus, $f_{\mathrm{raw}}$ is $C^1$-smooth on $[x_1,x_n]$, exactly interpolates the empirical data, and preserves monotonicity.

  \subsection{Asymptotic Exponential Extensions}
  To obtain a globally defined WE function, we extend $f_{\mathrm{raw}}$ outside its empirical domain using exponentially decaying tails that match both the value and derivative at the endpoints.

  Let
  \[
  x_L = x_1, \qquad f_L = y_1, \qquad d_L = m_1,
  \]
  \[
  x_R = x_n, \qquad f_R = y_n, \qquad d_R = m_n.
  \]

  \subsubsection{Left Tail: Approaching Zero}
  For $x < x_L$, we define
  \[
  f(x) = f_L \exp\!\big( k_L (x - x_L) \big),
  \qquad
  k_L = \frac{d_L}{f_L}.
  \]
  This ensures
  \[
  f(x_L^-) = f_L,
  \qquad
  f'(x_L^-) = d_L,
  \qquad
  \lim_{x\to -\infty} f(x) = 0.
  \]

  \subsubsection{Right Tail: Approaching One}
  For $x > x_R$, we define
  \[
  f(x)
  =
  1 - (1 - f_R) \exp\!\big( k_R (x_R - x) \big),
  \qquad
  k_R = \frac{d_R}{1 - f_R}.
  \]
  Thus,
  \[
  f(x_R^+) = f_R,
  \qquad
  f'(x_R^+) = d_R,
  \qquad
  \lim_{x\to +\infty} f(x) = 1.
  \]

  \subsection{Final Extended WE Function}
  Combining the interior spline and the asymptotic tails, the final win expectancy function is
  \[
  f(x) =
  \begin{cases}
  f_L \exp\!\big(k_L(x - x_L)\big), & x < x_L, \\[6pt]
  f_{\mathrm{raw}}(x), & x_L \le x \le x_R, \\[6pt]
  1 - (1-f_R)\exp\!\big(k_R(x_R - x)\big), & x > x_R.
  \end{cases}
  \]
  This function is continuous, continuously differentiable, strictly monotone, and exactly matches all empirical WE values. Moreover, it satisfies the natural asymptotic limits of win expectancy as score difference becomes extreme.

\end{appendices}

\end{document}